\documentclass[twocolumn,secnumarabic,amssymb, amsmath, nofootinbib,tightenlines,
nobibnotes, aps, prl,epsfig]{revtex4}
\usepackage{graphicx}
\usepackage{dcolumn}
\usepackage{bm}
\begin{document}
\preprint{APS/123-QED}
\title{Nonlinear Correction to the Longitudinal Structure Function at Small $x$}

\author{G.R.Boroun }
\altaffiliation{boroun@razi.ac.ir}
\affiliation{ Physics Department, Razi University, Kermanshah
67149, Iran}
\date{\today}

\begin{abstract}
We computed the longitudinal proton structure function $F_{L}$,
using the nonlinear Dokshitzer-Gribov-Lipatov-Altarelli-parisi
(NLDGLAP) evolution equation approach at small $x$. For the gluon
distribution, the nonlinear effects are related to the
longitudinal structure function. As, the very small $x$ behavior
of the gluon distribution is obtained by solving the Gribov,
Levin, Ryskin, Mueller and Qiu (GLR-MQ) evolution equation with
the nonlinear shadowing term incorporated. We show, the strong
rise that is corresponding to the linear QCD evolution equations,
can be tamed by screening effects. Consequently, the obtained
longitudinal structure function shows a tamed growth at small $x$.
We computed the predictions for all detail of the nonlinear
longitudinal structure function in the kinematic range where it
has been measured by $H1$ collaboration and compared with
computation Moch, Vermaseren and Vogt at the second order with
input data from MRST QCD fit.
\end{abstract}
\maketitle
\subsection{1 Introduction}
The measurement of the longitudinal structure function
$F_{L}(x,Q^{2})$ is of great theoretical importance, since it may
allow us to distinguish between different models describing the
QCD evolution at low-$x$. In deep- inelastic scattering (DIS), the
structure function measurements remain incomplete until the
longitudinal structure function $F_{L}$ is actually measured [1].
At small $x$ values, the dominant contribution to $F_{L}(x,Q^{2})$
comes from the gluon operators. Hence a measurement of
$F_{L}(x,Q^{2})$ can be used to extract the gluon structure
function and therefore the measurement of $F_{L}$ provides a
sensitive test of perturbative QCD [2-3]. As, at small $x$, the
longitudinal structure function can be related to the gluon and
sea- quark distribution. The behavior of the longitudinal
structure function at small $x$ is given by the gluon behavior.
The gluon behavior is observed that governs the physics of high
energy processes in QCD. HERA shows [4-9] that the gluon
distribution function has a steep behavior in the small x region
($10^{-2}{>}x{>}10^{-5}$). This steep behavior is well described
in the framework of the DGLAP [10-12] evolution equations.\\

In DIS at moderate values of $x$, the linear QCD evolution
equations lead to good description of this process. But at small
$x$, the problem is more complicated since recombination processes
between gluons in a dense system have to be taken into account.
This strong rise can eventually violate unitarity and so it has to
be tamed by screening effects. These screening effects are
provided by multiple gluon interaction which lead to the nonlinear
terms in the DGLAP equations. These nonlinear terms reduce the
growth of the gluon distribution in this kinematic region where
$\alpha_{s}$ is still small but the density of partons becomes so
large. Gribov, Levin, Ryskin, Mueller and Qiu (GLR-MQ)[13-14]
performed a detailed study of this region. They argued that the
physical processes of interaction and recombination of partons
become important in the parton cascade at a large value of the
parton density, and that these shadowing corrections could be
expressed in a new evolution equation (the GLR-MQ
equation)[13-14]. The main characteristic of this equation is that
it predicts a saturation of the gluon distribution at very small
$x$ [15-16]. This equation was based on two processes in a
parton cascade:\\
i)The emission induced by the QCD vertex $G{\rightarrow}G+G$ with
the probability which is proportional to $\alpha_{s}\rho$ where
$\rho(=\frac{xg(x,Q^{2})}{{\pi}R^{2}})$ is the density of gluon in
the transverse plane, ${\pi}R^{2}$ is the target area, and $R$ is the size of the target which the gluons populate;\\
ii)The annihilation of a gluon by the same vertex
$G+G{\rightarrow}G$ with the probability which is proportional to
$\frac{\alpha_{s}^{2}\rho^{2}}{Q^{2}}$, where $\alpha_{s}$ is
probability
of the processes.\\

Therefore, to obtain a precise evidence of the shadowing
correction in the HERA kinematic region, we consider the
longitudinal structure function that directly dependence on the
behavior of the gluon distribution. In this paper we estimate the
shadowing correction to the longitudinal structure function
behavior. We calculate this observable using the Altarelli-
Marinelli equation [17-18]. The longitudinal structure function
$F_{L}$, projected from the hadronic tensor by combination of the
metric and the spacelike momentum transferred by the virtual
photon $(g_{\mu\nu}-q_{\mu}q_{\nu}/q^{2})$. Indeed, the
longitudinal structure function is proportional to hadronic tensor
as follows:
$F_{L}(x,Q^{2})/x=\frac{8x^{2}}{Q^{2}}p_{\mu}p_{\nu}W_{\mu\nu}(x,Q^{2}),$
where $p^{\mu}(p^{\nu})$ is the hadron momentum and $W^{\mu\nu}$
is the hadronic tensor. In this relation we neglecting the hadron
mass. The basic hypothesis is that the total cross section of a
hadronic process can be written as the sum of the contributions of
each parton type (quarks, antiquarks, and gluons) carrying a
fraction of the hadronic total momentum. In the case of deep-
inelastic- scattering it reads:
\begin{equation}
d\sigma_{H}(p)=\sum_{i}{\int}dyd\hat{\sigma}_{i}(yp)\Pi_{i}^{0}(y),
\end{equation}
where $d\hat{\sigma}_{i}$ is the cross section corresponding to
the parton $i$ and $\Pi_{i}^{0}(y)$ is the probability of finding
this parton in the hadron target with the momentum fraction $y$.
Now, taking into account the kinematical constrains one gets the
relation between the hadronic and the partonic structure
functions:
\begin{eqnarray}
f_{j}(x,Q^{2})&=&\sum_{i}{\int}_{x}^{1}\frac{dy}{y}\textsf{f}_{j}(\frac{x}{y},Q^{2})\Pi_{i}^{0}(y)\nonumber\\
&&=\sum_{i}\textsf{f}_{j}{\otimes}\Pi_{i}^{0}(y)\hspace{0.3cm},j=2,L
\end{eqnarray}
where $\textsf{f}_{j}(x,Q^{2})=F_{j}(x,Q^{2})/x$. Equation (3)
expresses the hadronic structure functions as the convolution of
the partonic structure function, which are calculable in
perturbation theory, and the probability of finding a parton in
the hadron which is a nonperturbative function. So, in
correspondence with Eq.(3) one can write Eq.(1) as follows:
\begin{eqnarray}
F_{L}/x&=&\frac{\alpha_{s}}{4\pi}[\textsf{f}_{L,q}^{(1)}{\otimes}(q_{S}^{0}+q^{0}_{NS})+\textsf{f}_{L,G}^{(1)}{\otimes}g^{0}]
\end{eqnarray}
where $q^{0}_{S}$ and $q^{0}_{NS}$ are the singlet and nonsinglet
quark distribution. $\textsf{f}_{L,q}^{(1)}$ and
$\textsf{f}_{L,G}^{(1)}$ are the LO partonic longitudinal
structure function corresponding to quarks and gluons,
respectively [19-20]. At small $x$ the second term with the gluon
density is the dominant one. Here the representation for the gluon
distribution $G(x,Q^{2})=xg(x,Q^{2})$ is used, where $g(x,Q^{2})$
is the gluon density. After full agreement has been achieved, in
the form of the gluon kernel $K^{G}$, the standard collinear
factorization formula for the longitudinal structure function at
low $x$ reads:
 \begin{eqnarray}
F_{L}(x,Q^{2})=\int_{x}^{1}\frac{dy}{y}K^{G}(\frac{x}{y},Q^{2})G(y,Q^{2}).
\end{eqnarray}
where kernel $K^{G}$ is defined by:
\begin{eqnarray}
K^{G}(\frac{x}{y},Q^{2})=\frac{\alpha_{s}}{4\pi}[8(x/y)^{2}(1-x/y)][\sum_{i=1}^{N_{f}}e_{i}^{2}].
\end{eqnarray}
and $e_{i}$ are the quark charges.\\

One of the striking discoveries at HERA is the steep rise of the
gluon distribution function with decreasing $x$ value [6]. Indeed,
considering the HERA data, as is shown,
$G(x,Q^{2})=A_{g}x^{-\lambda_{g}(Q^{2})}$, where
$\lambda_{g}(Q^{2})$ is the Pomeron intercept mines one. As
$x{\rightarrow}0$ the value of the gluon density becomes so large
that the annihilation of gluons becomes important. So, this
singular behavior is tamed by the shadowing effects. The strategy
in this paper is based on the Regge- like behavior of the gluon
distribution function that tamed with the shadowing correction. We
assume this behavior as:
\begin{equation}
G^{sh}(x,Q^{2})=A_{g}x^{-\lambda^{sh}_{g}(Q^{2})}.
 \end{equation}
We note that at $x<x_{0}=10^{-2}$, shadowing gluon distribution
(sh.) and unshadowing gluon distribution (unsh.) behavior are
equal.  At $Q^{2}_{0}$ the small $x$ behavior of the shadowing
gluon distribution assumed to be[21-24]:
\begin{eqnarray}
G^{sh}(x,Q^{2}_{0})=G^{unsh}(x,Q^{2}_{0})[1+\theta(x_{0}-x)[G^{unsh}(x,Q^{2}_{0})\nonumber\\
-G^{unsh}(x_{0},Q^{2}_{0})]/xg_{sat}(x,Q^{2}_{0})]^{-1},
\end{eqnarray}
where
$xg_{sat}(x,Q^{2})=\frac{16R^{2}Q^{2}}{27\pi\alpha_{s}(Q^{2})}$ is
the value of the gluon which would saturate the unitarity limit in
the leading shadowing approximation. Based on this behavior, the
shadowing exponent of the gluon distribution can be determined as,
\begin{eqnarray}
\lambda_{g}^{sh}(Q^{2}_{0})&=&\lambda_{g}^{unsh}(Q^{2}_{0})+\frac{1}{Lnx}Ln[1+\theta(x_{0}-x){\times}\nonumber\\
&&[G^{unsh}(x,Q^{2}_{0})-G^{unsh}(x_{0},Q^{2}_{0})]{\times}\nonumber\\
&&\frac{27\pi^{2}}{4\beta_{0}R^{2}\Lambda^{2}t_{0}^{2}exp(t_{0})}]
\end{eqnarray}
here
$\alpha^{LO}_{s}(Q^{2})=\frac{4\pi}{\beta_{0}\ln(\frac{Q^{2}}{\Lambda^{2}})}$,
$\beta_{0}=\frac{1}{3}(33-2N_{f})$ and $N_{f}$ being the number of
active quark flavors ($N_{f}=4$), also
t=ln$(\frac{Q^{2}}{\Lambda^{2}})$,
$t_{0}$=ln$(\frac{Q_{0}^{2}}{\Lambda^{2}})$ (that $\Lambda$ is the
QCD cut- off parameter, i.e., $\Lambda\simeq0.2\hspace{0.1cm}GeV$
). The value of $R$ depends on how the gluon ladders couple to the
proton, or on how the gluons are distributed within the proton.
$R$ will be of the order of the proton radius
$(R\simeq5\hspace{0.1cm} GeV^{-1})$ if the gluons are spread
throughout the entire nucleon, or much smaller
$(R\simeq2\hspace{0.1cm} GeV^{-1})$ if gluons are concentrated in
hot- spot [25] within the proton. This equation (Eq.9) gives the
shadowing exponent of the shadowing gluon distribution function at
the scale $Q^{2}=Q^{2}_{0}$. In order to solve this equation [26]
we take $\lambda^{unsh}(Q_{0}^{2})$ with respect to
$G^{unsh}(x,Q_{0}^{2})$ that is the input unshadowing gluon
distribution that
take from QCD parametrisation.\\

Applying the dominant shadowing gluon distribution (i.e.Eq.7), in
order to calculate of the shadowing longitudinal structure
function at small $x$ to equation (5). After integration, we find
that:
\begin{equation}
F_{L}^{sh}(x,t)=\frac{20\alpha_{s}}{9\pi}
G^{sh}(x,t)Y_{1}(\lambda^{sh}_{g}(t)),
\end{equation}
where
\begin{eqnarray}
Y_{1}(\lambda^{sh}_{g}(t))=\hspace{5cm}\nonumber\\
\frac{(2+\lambda^{sh}_{g}(t))x^{3+\lambda^{sh}_{g}(t)}-(3+\lambda^{sh}_{g}(t))x^{2+\lambda^{sh}_{g}(t)}+1}{(2+\lambda^{sh}_{g}(t))(3+\lambda^{sh}_{g}(t))}.
\end{eqnarray}
The shadowing gluon distribution function should be defined in
this equation (GLR-MQ equation) as:
\begin{eqnarray}
\frac{dG^{sh}(x,Q^{2})}{dlnQ^{2}}&=&\frac{dG(x,Q^{2})}{dlnQ^{2}}|_{DGLAP}\nonumber\\
&&-\frac{81\alpha^{2}_{s}}{16R^{2}Q^{2}}\int^{1}_{x}\frac{dy}{y}G^{2}(y,Q^{2}),
\end{eqnarray}
where we used the modified gluon evolution equation  arise from
fusion of two gluon ladders. In this equation the first term is
the standard DGLAP result that is linear into the parton
distribution functions. Since we are interesting to evolution of
the longitudinal structure function with respect to nonlinear
corrections, we can easily perform this behavior using the
following equation:
\begin{eqnarray}
\frac{dF^{sh}_{L}(x,t)}{dt}=\frac{20}{9\pi}\frac{d\alpha}{dt}\int_{x}^{1}\frac{dy}{y}(\frac{x}{y})^{2}
(1-\frac{x}{y})G^{sh}(y,Q^{2})\nonumber\\
+\frac{20\alpha}{9\pi}\int_{x}^{1}\frac{dy}{y}(\frac{x}{y})^{2}
(1-\frac{x}{y})\frac{dG^{sh}(y,Q^{2})}{dt}.
\end{eqnarray}
where the derivative of the shadowing gluon distribution with
respect to $t$ is given by Eq.12. Since we have assumed that the
gluon have the Regge behavior at low $x$ as controlled by
shadowing corrections, we can easily solve this equation with
respect to this behavior. We find the derivative of shadowing
structure function with respect to $t$ as:
\begin{eqnarray}
\frac{dF^{sh}_{L}(x,t)}{dt}=(-\frac{20}{9t\pi}Y_{1}
+\frac{20\alpha^{2}}{3\pi^{2}}Y_{2})G^{sh}(x,t)\nonumber\\
-\frac{45\alpha^{3}}{8{\pi}R^{2}Q^{2}}Y_{3}G^{2sh}(x,t)
\end{eqnarray}
where
\begin{eqnarray}
Y_{2}=\frac{(1-x^{2+\lambda^{sh}_{g}(t)})}{\lambda^{sh}_{g}(t)(2+\lambda^{sh}_{g}(t))}-
\frac{(1-x^{3+\lambda^{sh}_{g}(t)})}{\lambda^{sh}_{g}(t)(3+\lambda^{sh}_{g}(t))}\\\nonumber
-\frac{x^{\lambda^{sh}_{g}(t)}(1-x^{2})}{2\lambda^{sh}_{g}(t)}+\frac{x^{\lambda^{sh}_{g}(t)}(1-x^{3})}{3\lambda^{sh}_{g}(t)}
\end{eqnarray}
and
\begin{eqnarray}
Y_{3}=\frac{(1-x^{2+2\lambda^{sh}_{g}(t)})}{\lambda^{sh}_{g}(t)(2+2\lambda^{sh}_{g}(t))}-
\frac{(1-x^{3+2\lambda^{sh}_{g}(t)})}{\lambda^{sh}_{g}(t)(3+2\lambda^{sh}_{g}(t))}\\\nonumber
-\frac{x^{2\lambda^{sh}_{g}(t)}(1-x^{2})}{2\lambda^{sh}_{g}(t)}+\frac{x^{3\lambda^{sh}_{g}(t)}(1-x^{3})}{3\lambda^{sh}_{g}(t)}
\end{eqnarray}

Therefore, the following equation is a formula to extracted the
shadowing longitudinal structure function, using the shadowing
gluon distribution exponent and the shadowing gluon distribution
determined in [26] at small $x$ as a function of $t$ value with
respect to the initial conditions at $t=t_{0}$,as
\begin{eqnarray}
F^{sh}_{L}(x,t)=\frac{e^{\int-Y_{4}(t')/t'dt'}}{\int{e^{\int-Y_{4}(t')dt'}}Y_{5}(t')/t'dt'+C}
\end{eqnarray}
where
\begin{eqnarray}
Y_{4}=1-\frac{12}{\beta_{0}}\frac{Y_{2}}{Y_{1}}
\end{eqnarray}
and
\begin{eqnarray}
Y_{5}=\frac{3645\pi^{2}}{800R^{2}Q^{2}\beta_{0}}\frac{Y_{3}}{Y_{1}^{2}}
\end{eqnarray}
in this equation $C$ is a constant and dependence to the initial
conditions at $t=t_{0}$ and $x=x_{0}$.  The results are shown in
Fig.1 for $Q^{2}=20\hspace{0.1cm}GeV^{2}$ at hot- spot point with
$R=2\hspace{0.1cm}GeV^{-1}$.\\

In Fig.1, the values of the nonlinear longitudinal structure
functions are compared with the experimental $H1$ data[6,27].
Also, we compared our results with
 predictions of $F_{L}$ up to NLO in perturbative QCD [28-32] that
 the input densities is given by MRST parameterizations [33]. The simple conclusions,
which could be obtained from this figure, is the following. First
of all, our results at hot- spot for  $Q^{2}$ constant give
 values comparable of the shadowing longitudinal structure functions that are comparing with
the experimental data. They grow both with the rapidity $1/x$.
Secondly, our data show that shadowing longitudinal structure
function increase as x decreases, that its corresponding with PQCD
fits at low x, but this behavior tamed with respect to nonlinear
terms at GLR-MQ equation.  Consequently, our results based on the
Regge- like behavior of the shadowing gluon distribution function
slower evolution of the shadowing longitudinal structure function
in the nonlinear case at small values of $x$. These effects show
that their strong rise are taming in accordance with unitarity of
the
description by the interactions between gluons.\\

In conclusion, in this paper we have obtained  the effects of
adding the nonlinear GLR-MQ corrections to the DGLAP evolution
equation and especially the shadowing effects to the longitudinal
structure function at low $x$. We saw that the gluon recombination
effects are expected to play an increasingly important role. These
effects that arise from fusion of the two gluon ladders, slow down
the evolution of the gluons from the standard DGLAP behavior. We
show that the obtained results for the shadowing longitudinal
structure function at small-$x$ have a power- like behavior. As
this growth tamed by the shadowing effects. This implies that the
$x$ dependence of the shadowing
 longitudinal structure function at low $x$ is consistent with a
 power law, $F^{sh}_{L}(x,Q^{2})=A_{L}x^{-\lambda^{sh}_{L}(Q^{2})}$, for fixed $Q^{2}$.
 This behavior is associated with the exchange of an object known
 as the hard Pomeron and also exponent of the
shadowing longitudinal structure function defined
as a polynomial function with respect to $lnQ^{2}$.\\

\textbf{References}\\
\hspace{2cm}1. A.Gonzalez-Arroyo, C.Lopez, and F.J.Yndurain, phys.lett.B\textbf{98}, 218(1981).\\
\hspace{2cm}2. A.M.Cooper- Sarkar, G.Inglman, K.R.Long, R.G.Roberts, and D.H.Saxon , Z.Phys.C\textbf{39}, 281(1988).\\
\hspace{2cm}3. R.G.Roberts, The structure of the proton, (Cambridge University Press 1990)Cambridge.\\
\hspace{2cm}4. S.Aid et.al, $H1$ collab. phys.Lett. {\bf B393}, 452-464 (1997).\\
\hspace{2cm}5. R.S.Thorne, phys.Lett. {\bf B418}, 371(1998).\\
\hspace{2cm}6. C.Adloff et.al, $H{1}$ Collab., Eur.Phys.J.C\textbf{21}, 33(2001).\\
\hspace{2cm}7. N.Gogitidze et.al, $H{1}$ Collab., J.Phys.G\textbf{28}, 751(2002).\\
\hspace{2cm}8. A.V.Kotikov and G.Parente,
JHEP\textbf{85},17(1997);
Mod.Phys.Lett.A\textbf{12}, 963(1997).\\
\hspace{2cm}9. C.Adloff,$H1$ collab. phys.Lett. {\bf B393}, 452(1997).\\
10. Yu.L.Dokshitzer, Sov.Phys.JETP {\textbf{46}},
641(1977).\\
11. G.Altarelli and G.Parisi, Nucl.Phys.B \textbf{126},
298(1977).\\
12. V.N.Gribov and L.N.Lipatov,
Sov.J.Nucl.Phys. \textbf{15}, 438(1972).\\
13. L.V.Gribov, E.M.Levin and M.G.Ryskin, Phys.Rep.\textbf{100},
 1(1983).\\
14. A.H.Mueller and J.Qiu, Nucl.Phys.B\textbf{268}, 427(1986).\\
15. A.L.A.Filho. M.B.Gay Ducati and V.P.Goncalves,
 Phys.Rev.D\textbf{59},054010(1999).\\
16. K.J.Eskola, et.al., Nucl.Phys.B660, 211(2003).\\
 17. G.Altarelli and G. Martinelli, Phys.Lett.B\textbf{76},89(1978).\\
 18. K.G.Biernat and A.M.Stasto,hep-ph/arXiv:0905.1321.\\
19. D.I.Kazakov, et.al., Phys.Rev.Lett. \textbf{65}, 1535(1990).\\
20. K.G.Biernat,hep-ph/arXiv:0207188. \\
21. A.D.Martin, W.J.Stirling, R.G.Roberts,  and R.S.Thorne,
Phys.Rev.D \textbf{47}, 867(1993).\\
22. J.Kwiecinski, A.D.Martin and P.J.Sutton, Phys.Rev.D
\textbf{44}, 2640(1991).\\
23. A.J.Askew, J.Kwiecinski, A.D.Martin and P.J.Sutton, Phys.Rev.D
\textbf{47}, 3775(1993).\\
24. J.Kwiecinski and A.M.Stasto, Phys.Rev.D \textbf{66},
014013(2002).\\
25. E.M.Levin and M.G.Ryskin, Phys.Rep.\textbf{189}, 267(1990).\\
26.G.R.Boroun, Eur. Phys.J.A42, 251(2009); Int.J.Mod.Phys.E18, No.1,1(2009).\\
27.H1 Collab., F.D.Aaron, et.al., Phys.Lett.B665, 139(2008).\\
28. A. Donnachie and P.V.Landshoff, Phys.Lett.B\textbf{533},
277(2002); Phys.Lett.B\textbf{550}, 160(2002).\\
29. J.R.Cudell, A.
Donnachie and P.V.Landshoff, Phys.Lett.B\textbf{448}, 281(1999).\\
30. P.V.Landshoff, hep-ph/0203084.\\
31. A.Vogt, S.Moch, J.A.M.Vermaseren, Nucl.Phys.B \textbf{691},
129(2004).\\
32. S.Moch, J.A.M.Vermaseren, A.vogt, Phys.Lett.B \textbf{606},
123(2005).\\
33. A.D.Martin, R.G.Roberts, W.J.Stirling,R.Thorne, Phys.Lett.B
\textbf{531}, 216(2001).

\subsection{Figure captions }
Fig 1:The values of the shadowing longitudinal structure function
at $Q^{2}=20\hspace{0.1cm} GeV^{2}$ with
    $R=2\hspace{0.1cm}GeV^{-1}$ (square) that accompanied with model error by solving the
GLR-MQ evolution equation
  that compared with H1 Collab. data
    (up and down triangle). The error on the  H1
 data is the total uncertainty of the determination of
 $F_{L}$ representing the statistical, the systematic and the model errors added in quadrature.
Circle data are the MVV prediction [31-33]. The solid line is the
NLO QCD fit to the H1 data for $y<0.35$ and
  $Q^{2}{\geq}3.5\hspace{0.1cm}GeV^{2}$. \\
\begin{figure}
\includegraphics[width=1\textwidth]{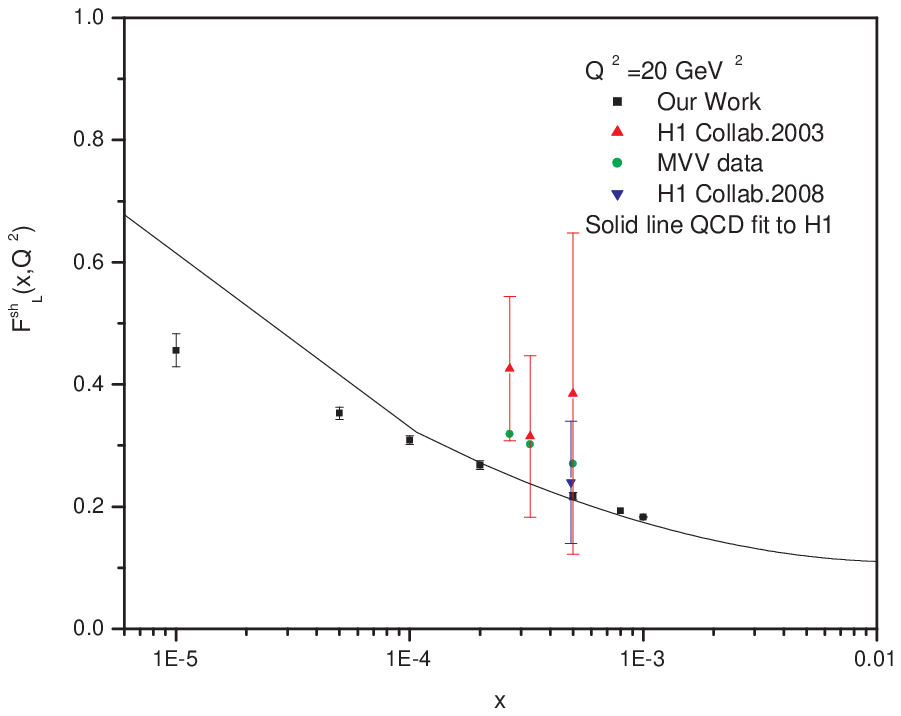}
\caption{} \label{Fig1}
\end{figure}
\end{document}